\documentclass[%
 reprint,
 aps,
]{revtex4-2}

\usepackage{graphicx}
\usepackage{bm}
\usepackage{mathtools}
\usepackage{subcaption}
\usepackage{placeins}

\begin{document}

\preprint{APS/123-QED}

\title{A cell-level model to predict the spatiotemporal dynamics of neurodegenerative disease}

\author{Shih-Huan Huang}
 \affiliation{Yusuf Hamied Department of Chemistry, University of Cambridge, Cambridge, UK.}
\author{Matthew W. Cotton}
 \affiliation{Yusuf Hamied Department of Chemistry, University of Cambridge, Cambridge, UK.}
\author{Tuomas P.J. Knowles}
 \affiliation{Yusuf Hamied Department of Chemistry, University of Cambridge, Cambridge, UK.}
  \altaffiliation{Cavendish laboratory, University of Cambridge, Cambridge, UK.}
 
\author{David Klenerman}
\email{dk10012@cam.ac.uk}
 \affiliation{Yusuf Hamied Department of Chemistry, University of Cambridge, Cambridge, UK.}
 
\author{Georg Meisl}%
\email{gm373@cam.ac.uk}
 \affiliation{Yusuf Hamied Department of Chemistry, University of Cambridge, Cambridge, UK.}

\begin{abstract}

A central challenge in modeling neurodegenerative diseases is connecting cellular-level mechanisms to tissue-level pathology, in particular to determine whether pathology is driven primarily by cell-autonomous triggers or by propagation from cells that are already in a pathological, runaway aggregation state. To bridge this gap, we here develop a bottom-up physical model that explicitly incorporates these two fundamental cell-level drivers of protein aggregation dynamics. We show that our model naturally explains the characteristic long, slow development of pathology followed by a rapid acceleration, a hallmark of many neurodegenerative diseases. Furthermore, the model reveals the existence of a critical switch point at which the system's dynamics transition from being dominated by slow, spontaneous formation of diseased cells to being driven by fast propagation. This framework provides a robust physical foundation for interpreting pathological data and offers a method to predict which class of therapeutic strategies is best matched to the underlying drivers of a specific disease. 
\end{abstract}

\maketitle

\section{\label{sec:level1.1}Introduction}
Current understanding of neurodegenerative diseases is limited by the challenge of connecting the microscopic mechanisms of protein aggregation to the macroscopic patterns of pathology observed in patients. While the general link between protein aggregation and disease is well-established~\cite{Dobson2017}, the physical principles governing the emergence of disease-specific patterns, such as the distinct spatial patterns or the characteristic long, slow initial phase followed by rapid progression~\cite{Jack2010}, remain poorly understood, hindering the development of rationally designed therapies~\cite{Meisl_Review2020}. 

\begin{figure}
  \includegraphics[width=1.0\linewidth]{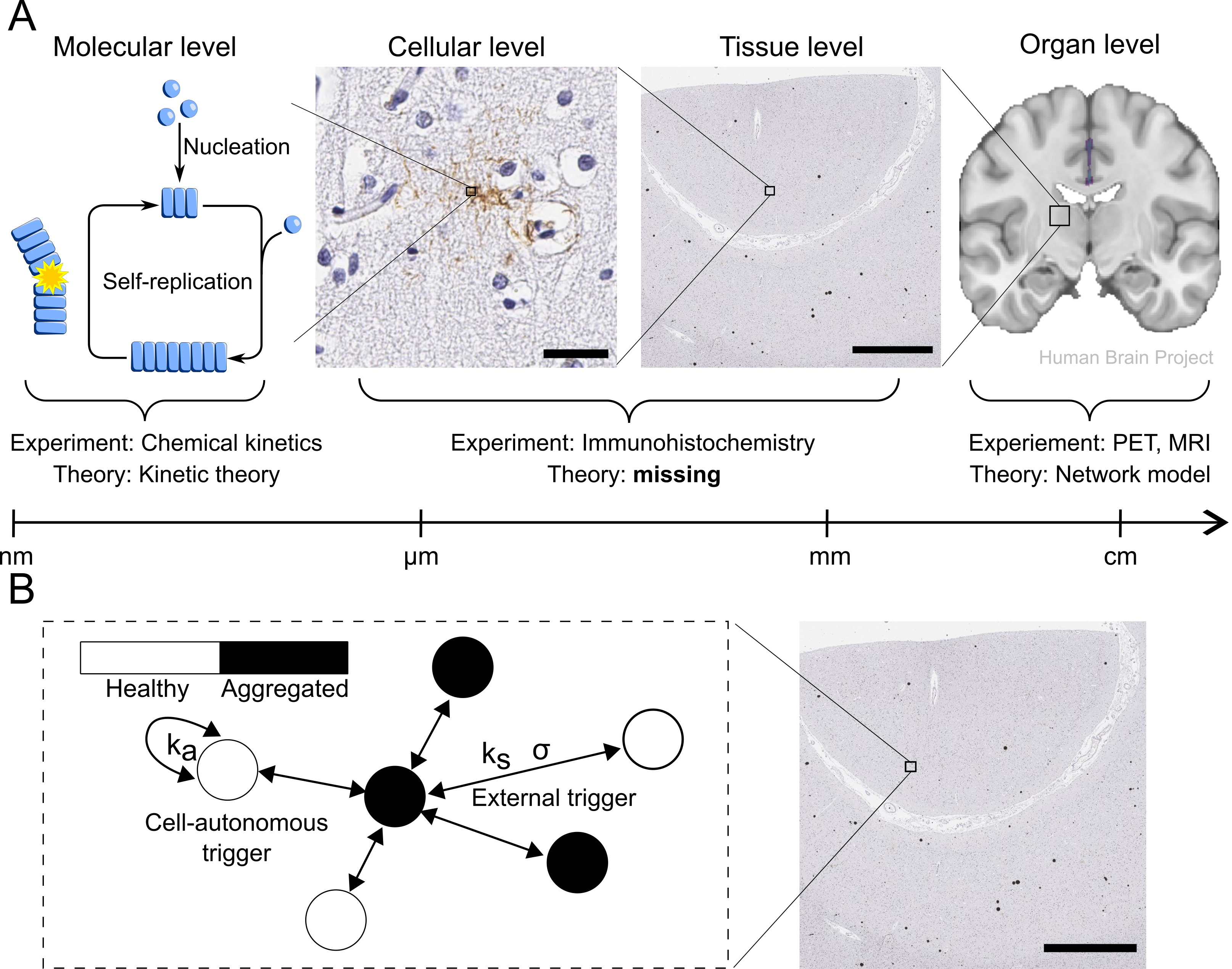}
  \caption{\textbf{Model description.}}
  (A) A central problem in neurodegenerative research is the lack of theory that can describe different systems across a wide range of temporal and spatial scales.  Scale bar: cellular level: 30 $\mu m$; tissue level: 2 mm. The brain image is adapted from Human Brain Project~\cite{Amunts2021}. (B) In our model, the fundamental unit is a cell. Aggregation can begin autonomously within a cell at a rate $k_a$ (measured in inverse time). Additionally, cells that have already aggregated can trigger aggregation in neighboring cells. This interaction depends on the coupling strength or rate $k_s$ (also measured in inverse time) and a length-dependent factor characterized by $\sigma$ (with units of length). 
  \label{fig:problem}
\end{figure}

This difficulty stems from the fundamental multiscale nature of the problem (Fig. \ref{fig:problem}A). Current kinetic theories describe protein filament assembly at the molecular level~\cite{Knowles2009,Cohen2011,Meisl2016_amylofit} and network models capture phenomenological spreading at the organ level~\cite{Vogel2023,Fornari2020}, but the crucial mesoscopic, or tissue, scale is left without a first-principles framework. There is thus currently no theory that connects the behaviour of individual cells to the emergent aggregation patterns observed in patient tissue. 


This theoretical gap is particularly problematic as it hinders our ability to resolve a central mechanistic question: does pathology arise primarily from cell-autonomous triggers within vulnerable cells, or is the major driver a spreading mechanism from cells that are already in a pathological, runaway aggregation state~\cite{Mudher2017, Clavaguera2009}? The challenge in distinguishing the contributions of these two fundamental mechanistic pathways from pathological data has been a major barrier to progress in the field~\cite{Meisl2021_tau_paper} as effective therapies should target fundamentally different processes in the two scenarios. 


To bridge this gap, we develop a bottom-up physical multiscale model connecting molecular, cellular, and tissue levels. Our framework explicitly incorporates both cell-autonomous triggers and cell-to-cell propagation to simulate the stochastic evolution of aggregation at the single-cell level. We showcase the varied behaviour this minimal model gives rise to in different regimes and discuss the mathematical relationship to reaction diffusion equations and epidemiological models of the spread of disease in a susceptible population. By deriving analytical expressions for key spatial statistics, we provide a quantitative toolkit to infer the dominant aggregation mechanism from static patterns of aggregates, culminating in the derivation of a critical switch fraction that predicts when the system transitions from being cell-autonomous to propagation-dominated.

\section{A cell-level model for neurodegenerative diseases}
\label{sec:model}

To bridge the gap between molecular events and tissue-level pathology, we developed a biophysical model to describe the spread of disease within a tissue. We model the tissue as an ensemble of discrete cells, the fundamental unit of biological organization~\cite{sequence_consequence}. Each cell $i$ is treated as a point-like compartment at position $\mathbf{x}_i$ and is described by a single binary state variable, $\phi_i$. This variable denotes whether the cell is `healthy' $(\phi_i=0)$ or has transitioned to a state of pathological protein aggregation $(\phi_i=1)$. 
We simulate the evolution of tissue by prescribing $\phi$ for every cell at some initial time and proceed in discrete time steps, $\Delta t$, during which a healthy cell can transition to the aggregated state through two distinct mechanistic pathways. 

The first pathway, the \emph{cell-autonomous triggers}, represents the spontaneous formation of protein aggregates within a single cell, independent of its neighbours. Depending on the disease, this cell-autonomous transition, could be caused by a stochastic nucleation event, or via overwhelming of the cellular aggregate removal mechanisms~\cite{Meisl2024}.  The effect of rare nucleation events has been well established in vitro~\cite{Knowles2011} and in vivo~\cite{Sinnige2021}, and is likely the key driver of sporadic prion disease.
By contrast, an overwhelming of aggregate removal mechanisms is a more likely scenario in common neurodegenerative disorders such as Alzheimer's disease. This process may be driven by factors such as age-related accumulation of somatic mutations in neurons~\cite{Bae2022}, or other stressors leading to a temporary failure of protein homeostasis~\cite{Labbadia2015} and ultimately runaway aggregation~\cite{Meisl_clearance}. We coarse grain these distinct processes into a single time-independent cell-autonomous rate, $k_a$. The probability $p_{a,i}$ that any given healthy cell $i$ spontaneously transitions to the aggregated state in a small time interval $\Delta t$ is therefore given by: 
$$p_{a,i}=1-\exp(-k_a\Delta t)\approx k_a\Delta t$$
for the short times $\Delta t$.

The second pathway for a cell to transition to the aggregated state is via \emph{cell-to-cell propagation}, which models the spread of pathology from an already aggregated cell to its healthy neighbours. This can represent various biological processes, including the direct transport of pathogenic protein seeds between cells or the triggering of inflammatory responses~\cite{Weickenmeier2018,tau_propogate_2009}, which in turn can compromise the protein quality control systems of otherwise healthy cells, increasing their vulnerability to aggregation. We model this coupling as a radially isotropic interaction function that decays with distance. Specifically, we use a Gaussian kernel for the majority of this work, as it represents a simple, canonical model for a diffusive process, but we also show that the conclusions are robust regardless of the choice of the specific interaction function. The probability, $p_{s,i}$, that a healthy cell $i$ is triggered by its aggregated neighbours during $\Delta t$ is then
$$p_{s,i} = 1-e^{-\lambda_i\Delta t},$$
where $\lambda_i$ is the local `cell-to-cell coupling rate':
$$\lambda_i=\frac{k_s\Sigma_j\left(\phi_j e^{-d_{ij}^2/2\sigma^2}\right)}{2\pi\sigma^2}.$$
Here, $d_{ij}$ is the distance between cell $i$ and cell $j$, $k_s$ is the spatial coupling strength (in units of time$^{-1}$), and $\sigma$ is the characteristic length scale of the interaction. The sum runs over all cells, but the non-zero contributions only come from the aggregated cells. In this work, we have also chosen a two dimensional system for convenience, but since diffusion scales with $\sqrt{t}$, independent of dimensionality, the results, such as the emergence of distinct regimes, the existence of a switch fraction, can trivially be extended to three dimensions.

We evaluate the dynamics of the system numerically using a discrete-time Markov chain: the state of every cell is updated at each time step  $\Delta t$ (see SI for validation of convergence Fig.~\ref{fig:SIFig1_time_conv}A and B). A healthy cell transitions to the aggregated state if it is triggered by either the cell-autonomous pathway (with probability $p_{a,i}$) or the cell-to-cell propagation pathway (with probability $p_{s,i}$), which are treated as independent stochastic events. We also define the two-dimensional density of cells as $\rho$, in units of distance$^{-2}$. In this work, we set the density to be constant to compare more easily with analytical results; however, this is not required by our simulation framework.

\section{Results}

To quantitatively characterize the spatial patterns of aggregation produced by our model, we employ three distinct measures, each sensitive to structure on a different length scale. The \emph{total fraction of aggregated cells} provides a macroscopic view of disease progression, analogous to readouts from bulk tissue homogenates~\cite{Böken2024}. For understanding the emerging spatial patterns,  we use two measures: The \emph{Nearest Neighbour Distance (NND) distribution} is the probability distribution of distances from each aggregated cell to its nearest non-self aggregated cell. The \emph{Radial Distribution Function (RDF)}, $g_{norm}(r)$, is here defined as the average fraction of cells that are aggregated a distance $r$ away from an aggregated cell. Mathematically, $g_{norm}(r)=\langle dn_r\rangle/\langle dn_c \rangle$ where $dn_{r}$ and $dn_c$ are the number of aggregated cells and the number of total cells, respectively, in an annulus of radius $r$ and infinitesimal width, $dr$, around each aggregated cell. Note that this also captures the overall degree of aggregation in its magnitude and thus differs in its normalisation from the standard definition of the radial distribution function in the statistical mechanics literature, where a value of 1 corresponds to a random arrangement at the average density. Together, these measures provide a comprehensive fingerprint of the aggregation pattern that can be directly compared to experimental data~\cite{Huang2024}.

\subsection{Spontaneous Aggregation ($k_s=0$) Results in Random Spatial Patterns}
We initially consider the simplest limiting case where aggregation is driven exclusively by cell-autonomous triggers ($k_s=0$). This scenario corresponds to a hypothetical disease process where cells are entirely independent, and pathology cannot spread between them. The aggregation of polyglutamine in a \textit{C. elegans} model of Huntington's disease is an example of such behaviour~\cite{Sinnige2021}. 

In this  limiting case, we predict three signatures: First, the total fraction of aggregated cells, $n$, increases as healthy cells transition to the aggregated state and is captured well by a simple analytical model of independent events~\cite{Sinnige2021}:
\begin{equation}
    n(t) = 1-e^{-k_{a}t},
    \label{eq:prim}
\end{equation}
where $k_a$ is the rate constant of the cell-autonomous process (see Appendix \ref{(SI)primary_frac_aggregated} for complete derivation).
Second, due to our specific normalisation, the RDF should be a constant with a value equal to the total fraction of aggregated cells (see Appendix.~\ref{Apdix:RDF_random_case} for full derivation), and, third, the NND distribution should follow the analytical form for a Poisson point process: 
\begin{equation}
\begin{aligned}
P(\text{nearest neighbor distance at } r\rightarrow r+dr)\\=  2\pi rDe^{-\pi r^2D}dr,
\label{eq:weakpsatialNDDdist}
\end{aligned}
\end{equation}
where $D$ is the density of aggregated cells (see Appendix \ref{(Apdix)primary_NND} for complete derivation). Here we have $D=\rho n$, where $\rho$ is the density of cells.

\begin{figure}
  \includegraphics[width=1\linewidth]{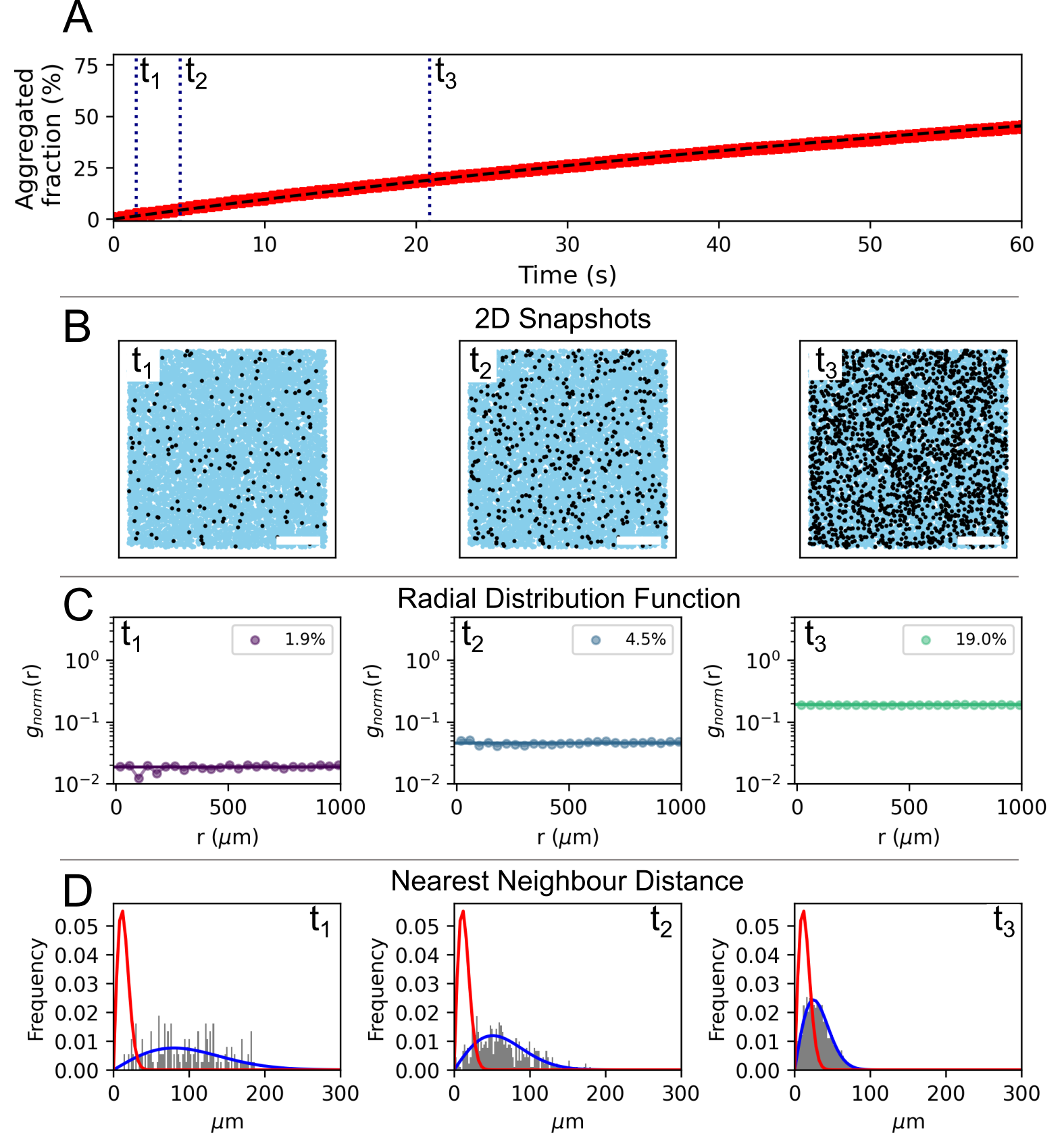}
\caption{\textbf{No spatial coupling case. }}(A) Temporal evolution of total aggregated fraction. The red dots are simulated points. The black dash line is the developed analytical theory (Eq.~\ref{eq:prim}). (B) Snapshots of simulations of no spatial coupling case at fraction aggregated = 2\%, 5\%, 20\%. Light blue points are non-aggregated cells, black points are aggregated cells. (C) RDF of (B) (D) Corresponding NND distribution of (B). Scale bar on (B): 500 \textmu m. Simulation conditions: total cell number = 10002, $k_a$ = 0.01/s, $k_s$ = 0/s, $\Delta t$ = 0.1 s, $\rho$ = 1341.76/$mm^2$, initial aggregated cell number = 0.
  \label{fig:primary}
\end{figure}

As shown in the simulation snapshots (Fig.~\ref{fig:primary}B), this regime produces a random, spatially uncorrelated scattering of aggregated cells across the tissue at all time points. This visual observation is confirmed quantitatively by the RDF (Fig.~\ref{fig:primary}C), as $g_{norm}(r)$ is constant and is equal to the total fraction of aggregated cells. In addition, a good match between simulated total cell fraction aggregated and the analytical expression can be seen in Fig.~\ref{fig:primary}A. Finally, the simulated NND distribution agrees well with the analytical expression in Eq. \eqref{eq:weakpsatialNDDdist} (Fig.~\ref{fig:primary}D). Note that at early times, the distribution appears noisy due to the low number of aggregated cells.
\begin{figure}
    \centering
    \includegraphics[width=1.0\linewidth]{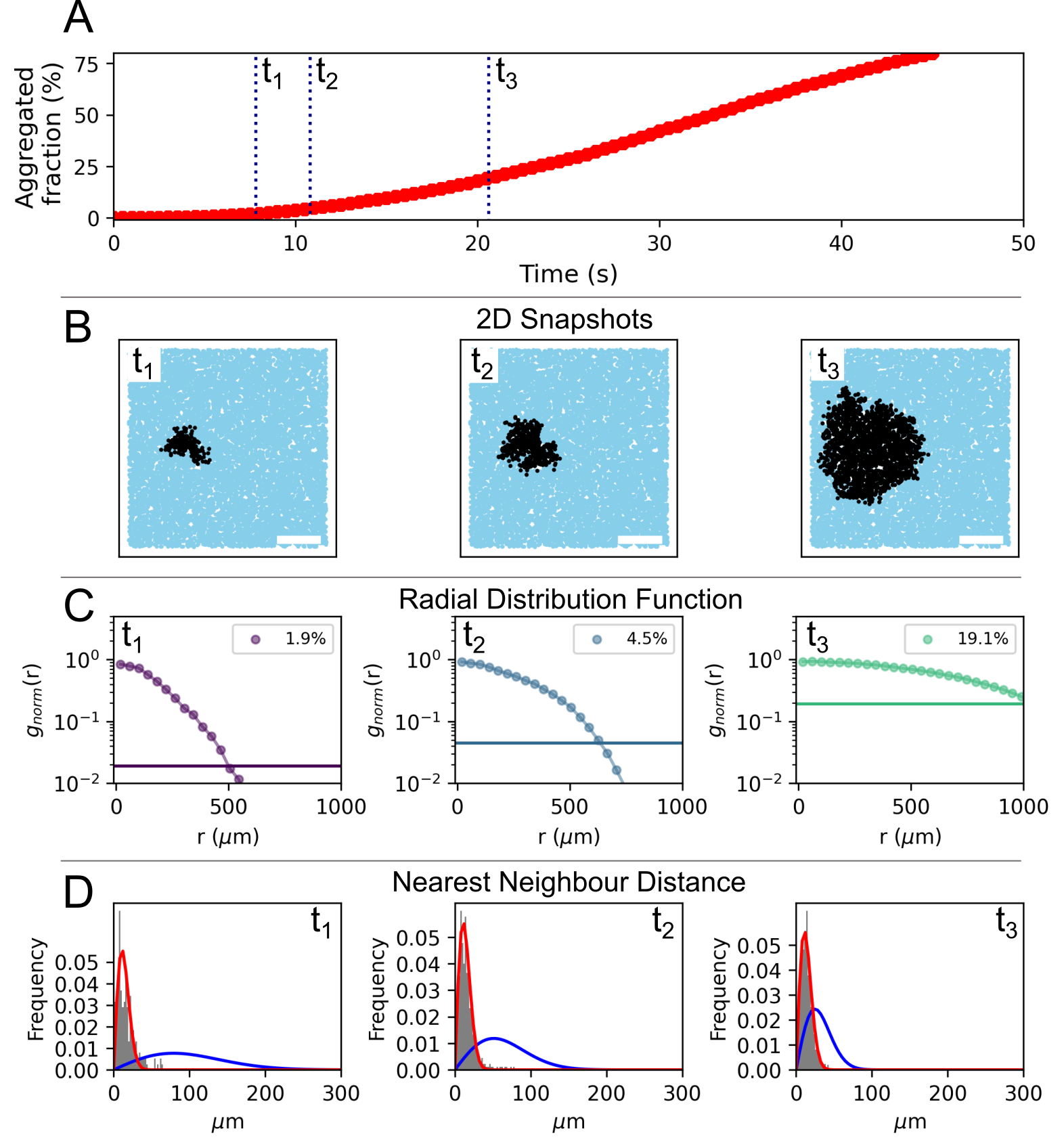}
    \caption{\textbf{Short-range coupling case. }}(A) Temporal evolution of total aggregated fraction. The red dots are simulated points. (B) Snapshots of simulations of the strong spatial coupling case at fraction aggregated = 2\%, 5\%, 20\%. (C) RDF of (B) (D). Corresponding NND distribution of (B). Scale bar on (B): 500 \textmu m. Simulation conditions: total cell number = 10000, $k_a$ = 0/s, $k_s$ = 1.0/s, $\sigma$ = 40$\mu m$, coupling kernel: Gaussian, $\Delta t$ = 0.1 s, $\rho$ = 1341.76/$mm^2$, initial aggregated cell number = 1.
    \label{fig:strong_coupling}
\end{figure}


\subsection{Propagating-Only Aggregation ($k_a\rightarrow 0$) Leads to Growing, Spatially Correlated Clusters}
In this limiting case, aggregation is driven solely by cell-to-cell propagation from an initial seed ($k_a\approx 0$). This case models a scenario where the disease spreads like an infection from a single starting point, a key feature of prion-like dynamics. The nature of this propagation depends critically on the interaction distance, $\sigma$, relative to the cell spacing, $\rho^{-1/2}$. We therefore analysed the behaviour in two distinct regimes: the short-range and long-range coupling limits.

\subsubsection{Short-Range Coupling Generates Dense Clusters}
First, we examined the short-range regime, where the interaction distance is comparable to the cell spacing $\sigma\sim\rho^{-1/2}$, modelling direct neighbour transmission. In this case, the simulations show the formation of a single cluster of aggregated cells, with a sharp boundary, that grows outwards as neighbouring cells are sequentially triggered (Fig.~ \ref{fig:strong_coupling}B). This dense region of aggregated cells creates two clear quantitative signatures. First, the NND distribution is sharply peaked at the mean intercellular distance and determined by the underlying density of all cells, $\rho$, not the number of aggregates (Fig.~ \ref{fig:strong_coupling}D, Eq.~\ref{eq:NNDD_short_coupling})
\begin{equation}
\begin{aligned}
P(\text{nearest neighbor distance at} r\rightarrow r+dr)\\=  2\pi r\rho e^{-\pi r^2\rho}dr,
\end{aligned}
\label{eq:NNDD_short_coupling}
\end{equation}

Second, the RDF shows a strong, sharp boundary, due to the cells forming a dense aggregated cluster (Fig.~ \ref{fig:strong_coupling}C). In this short-range coupling limit, when the separation of cells is greater than the characteristic length scale of the interaction, the exact shape of the aggregation front depends on the underlying lattice of cells and is a direct feature of the spatial discretisation. This aggregation pattern is therefore not not well-described by a simple continuous model.

\subsubsection{Long-Range Coupling Gives Rise to Continuous Traveling Waves}
\begin{figure}
  \includegraphics[width=\linewidth]{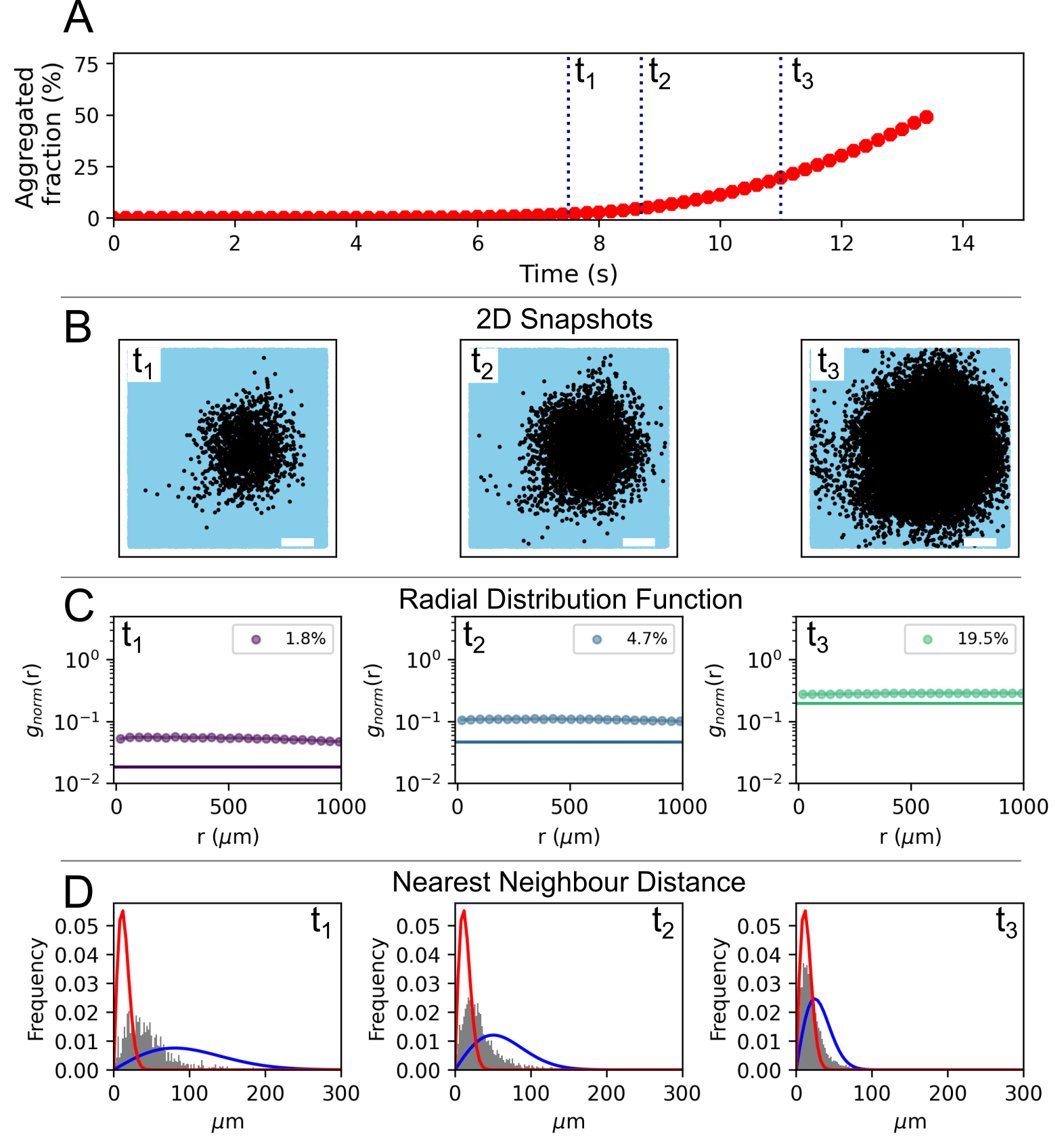}
  \caption{\textbf{Long range spatial coupling}.(A) Temporal evolution of total aggregated fraction. The red dots are simulated points. (B) Snapshots of simulations of the strong spatial coupling case at fraction aggregated = 2\%, 5\%, 20\%. (C) Radial distribution function of (B). (D) Corresponding nearest neighbour distance distribution of (B). Scale bar on (B): 1000 \textmu m. Simulation conditions: total cell number = 90000, $k_a$ = 0/s, $k_s$ = 1.0/s, $\sigma$ = 400$\mu m$, coupling kernel: Gaussian, $\Delta t$ = 0.1 s, $\rho$ = 1341.76/$mm^2$, initial aggregated cell number = 1.}
  \label{fig:long_range_coupling}
\end{figure}

We then investigated the long-range coupling regime, where the interaction distance is much larger than the cell spacing ($\sigma\gg \rho^{-1/2}$). This corresponds to propagation beyond direct neighbours and in this limit we expect a continuum approximation to be valid. Here, the aggregation front is no longer sharp, but rather a diffuse boundary, which leads to a broad correlation peak in the RDF (see Fig.~\ref{fig:long_range_coupling} where $\sigma\rho^{-1/2}\approx15$, compared to $\sigma\rho^{-1/2}\approx1.5$ in Fig. \ref{fig:strong_coupling}).

\begin{figure*}[t]
    \centering
    \includegraphics[width=\textwidth]{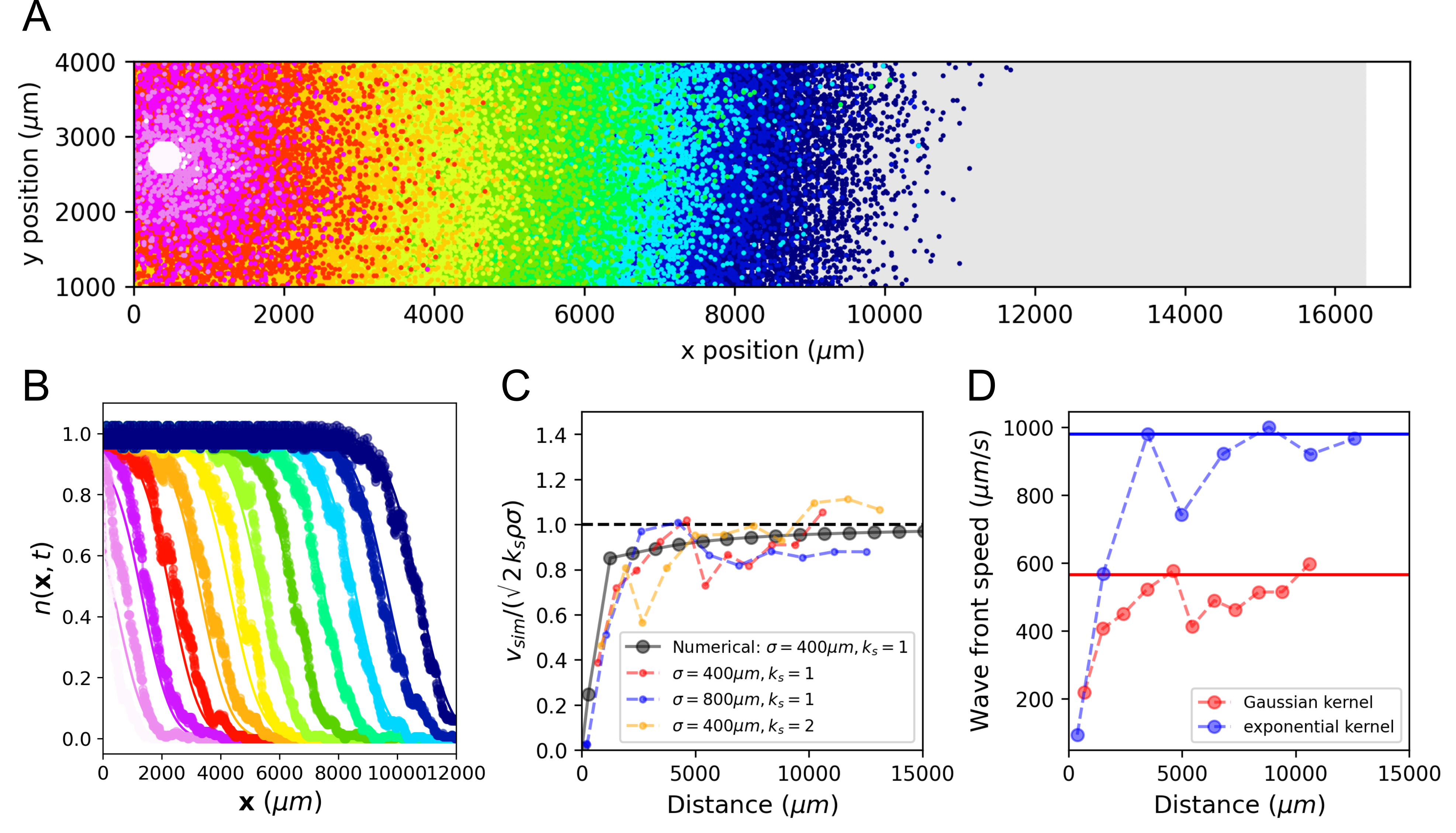}
    \caption{\textbf{Wavefront development in long-range coupling cases.}} (A) 2D snapshots of wavefront development in a strong spatial coupling case over a total of 20 sec. The colour represents time, progressing from early (pink) to late (blue). Simulation conditions: total cell number = 120000, $k_a$ = 0/s, $k_s$ = 1.0/s, $\sigma$ = 400$\mu m$, coupling kernel: Gaussian, $\Delta t$ = 0.1 s, $\rho$ = 1341.76/$mm^2$, with an initial condition of a circle of 166 aggregated cells centered at (409.5\textmu m, 2730\textmu m) . (B) Profiles of the wavefront from simulations in (A) compared to numerical integration of Eq.~\ref{pde_coupling}. For direct comparison, the simulation profiles have been spatially aligned such that the center of the initial seed corresponds to the origin ($x=0$). The numerical integration was performed using parameters matched to the simulation ($k_s$, $\sigma$, $\rho$), with no-flux boundary conditions and a step-function initial condition representing a fully aggregated region at the origin. The colour scheme is the same as panel A, with profiles shown at regular time intervals of 2 s.  (C) Wave speed comparison for different conditions. Simulation conditions are the same as Fig.~\ref{fig:wavefront_analysis}A except for changing  $k_s$ and $\sigma$. The different conditions are all normalised to the Fisher-KPP propagation speed (dashed line), which they approach at later times. (D) Wave speed comparison between Gaussian kernel and exponential kernel. Simulation conditions for exponential kernel: the same as Fig.~\ref{fig:wavefront_analysis}A with the coupling kernel being exponential decay. 
    \label{fig:wavefront_analysis}
\end{figure*}

When the length-scale of interactions is significantly longer than the length-scale of discretisation, a continuous model can be used to approximate the dynamics and derive closed form expressions that describe the evolution of the aggregation patterns. We describe the fraction of aggregated cells in a small volume at position $\mathbf{x}$ at time $t$ as $n(\mathbf{x}, t)$, so that the density of aggregated cells is $\rho(\mathbf{x}) n(\mathbf{x}, t)$. In this continuum limit, the cell-to-cell coupling rate at position $\mathbf{x}$ to aggregate is
\begin{equation}
    \tilde{\lambda}(\mathbf{x}, t)=\frac{k_s}{2\pi\sigma^2}\int_{\Omega}\rho(\mathbf{x'}) n(\mathbf{x'}, t)e^{\frac{-(\mathbf{x}-\mathbf{x'})^2}{2\sigma^2}}d^{2}\mathbf{x'},
    \label{eq:lamCont}
\end{equation}
where the integration is over the entire tissue, $\Omega$, and for the uniform density we have $\rho(\mathbf{x})=\rho$. Using a saddle point approximation,  $n(\mathbf{x'}, t)=n(\mathbf{x}, t)+\nabla n(\mathbf{x}, t)^{T}(\mathbf{x}-\mathbf{x'})+(1/2)(\mathbf{x}-\mathbf{x'})^{T}\mathrm{H}_{n}(\mathbf{x}, t)(\mathbf{x}-\mathbf{x'}) + \cdots$ where $\mathrm{H}_{n}(\mathbf{x}, t)$ is the Hessian matrix of $n$ evaluated at $(\mathbf{x}, t)$, yielding
\begin{equation}
    \tilde{\lambda}(\mathbf{x}, t) \approx k_s \rho n(\mathbf{x}, t) +\frac{1}{2} k_s \rho \sigma^{2}\nabla^{2}n(\mathbf{x}, t).
    \label{eq:lamExpand}
\end{equation}
At every point, only the non-aggregated cells can transition to the aggreagted state, so the fraction of aggregated cells will increase at a rate proportional to the number of non-aggregated cells giving 
\begin{equation}
\partial n(\mathbf{x}, t)/\partial t = \tilde{\lambda}(\mathbf{x}, t)(1-n(\mathbf{x}, t)).    
\label{integrodiffeq}
\end{equation}
Combining with Eq.~\eqref{eq:lamExpand}, we obtain a partial differential equation for the fraction of aggregated cells,
\begin{equation}
    \frac{\partial n(\mathbf{x}, t)}{\partial t} = \alpha(1-n(\mathbf{x}, t)) n(\mathbf{x}, t) + \beta (1-n(\mathbf{x}, t))\nabla^{2}n(\mathbf{x}, t),
\label{pde_coupling}
\end{equation}
with $\alpha=k_s \rho$ and $\beta=\frac{1}{2}k_s \rho \sigma^{2}$. We use numerical integration to obtain the solution to this nonlinear partial differential equation. As shown in Fig.~\ref{fig:wavefront_analysis}B, there is a close match between our discrete model and the numerical integrated results of Eq.~\eqref{pde_coupling}.

Eq.~\ref{pde_coupling} has the form of a Fisher-KPP reaction-diffusion equation~\cite{Fisher1937}, with an additional $(1-n)$ factor for the spatial term. For low densities where $1-n\approx1$, we can therefore obtain the limiting travelling wave speed as the speed of the Fisher wave
\begin{equation}
    v\geq\sqrt{2}k_s\rho\sigma.
    \label{eq:Fisher_KPP_speed}
\end{equation}
This analytical form for the wave speed still matches well our numerically integrated results of the travelling wave speed, as demonstrated in Fig.~\ref{fig:wavefront_analysis}C. A key observation is that, comparing the parameters in our model to those of Fisher-KPP, both the parameter for the local reaction, $\alpha$, as well as the parameter for `diffusion', $\beta$, depend on the rate or cell-autonomous triggers, $k_s$. In other words, the local reaction term is governed by the same parameter as the longer range interactions. 

Note that while deriving Eq.~\ref{pde_coupling}, we assume a Gaussian decay coupling, the results can be generalised for any symmetric, sufficiently rapidly decaying distribution. For example, changing the coupling term to an exponential decay yields
\begin{equation}
    \tilde{\lambda}(\mathbf{x}, t)=\frac{k_s}{2\pi\sigma^2}\int_{\Omega}\rho(\mathbf{x'}) n(\mathbf{x'}, t)e^{\frac{-|\mathbf{x}-\mathbf{x'}|}{\sigma}}d^{2}\mathbf{x'}.
    \label{eq:lamCont_exponential_decay}
\end{equation}
Using the same approach of Taylor expansion (Eq.~\ref{eq:lamExpand}) and growth dynamics (Eq.~\ref{integrodiffeq}), we can arrive at a partial differential equation of similar form,
\begin{equation}
    \frac{\partial n(\mathbf{x}, t)}{\partial t} = \alpha(1-n(\mathbf{x}, t)) n(\mathbf{x}, t) + \beta '(1-n(\mathbf{x}, t))\nabla^{2}n(\mathbf{x}, t),
\label{pde_coupling_exponential_decay}
\end{equation}
with $\alpha=k_s \rho$ and $\beta'=\frac{3}{2}k_s \rho \sigma^{2}$, the only difference being an additional factor of 3 in the definition of $\beta'$. Similarly, the corresponding Fisher travelling wave speed for the exponential decay kernel
\begin{equation}
    v=\sqrt{6}k_s\rho\sigma.
    \label{eq:Fisher_KPP_speed_exponential_decay}
\end{equation}
The analytical form for the wave speed again matches well our numerically integrated results of the travelling wave speed (Fig.~\ref{fig:wavefront_analysis}D), demonstrating ropbustness of the bahviour to the specific choice of the coupling function. 

It is interesting to note that our models resemble those used in epidemiology to model the spreading of pathology between individuals. When assuming there are no recovered or deceased individuals, the description for infectious disease follow similar dynamics to our systems in the propagation-only limit  (see ~\cite{Bartlett1957} and~\cite{Kendall1965}). Accordingly, the above derivation of a continuous model is similar to the integro-differential equation (Eq.~\ref{integrodiffeq}) used in previous work in epidemiology~\cite{MEDLOCK2003}. Similar reaction-diffusion frameworks have also been developed to model the spatial propagation of protein polymerisation in a homogeneous solution in vitro, where a Fisher wave emerges in the diffusion-dominated limit~\cite{Cohen2014PRL}.  
\begin{figure}
  \includegraphics[width=1\linewidth]{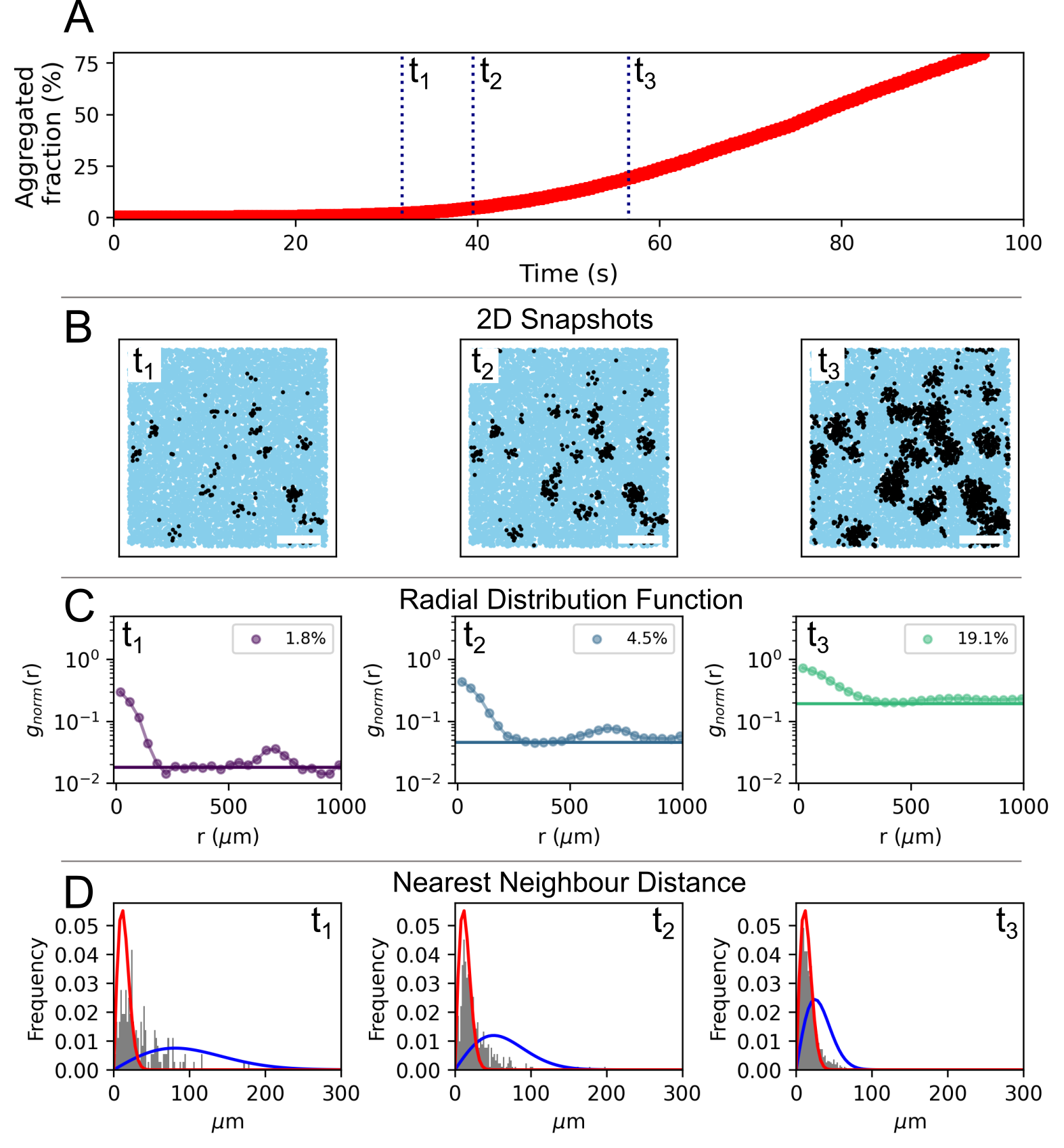}
\caption{\textbf{Intermediate regime - short range coupling}.} (A) Temporal evolution of total aggregated fraction. The red dots are simulated points. (B) Snapshots of simulations of weak spatial coupling case at fraction aggregated = 2\%, 5\%, 20\%. (C) RDF of (B) (D) Corresponding NND distribution of (B). Scale bar on (B): 500 \textmu m. Simulation conditions: total cell number = 10002, $k_a$ = 0.0001/s, $k_s$ = 1.0/s, $\sigma$ = 40$\mu m$, coupling kernel: Gaussian, $\Delta t$ = 0.1 s, $\rho$ = 1341.76/$mm^2$, initial aggregated cell number = 0.
  \label{fig:intermediate_regime}
\end{figure}
\begin{figure}
  \includegraphics[width=1.0\linewidth]{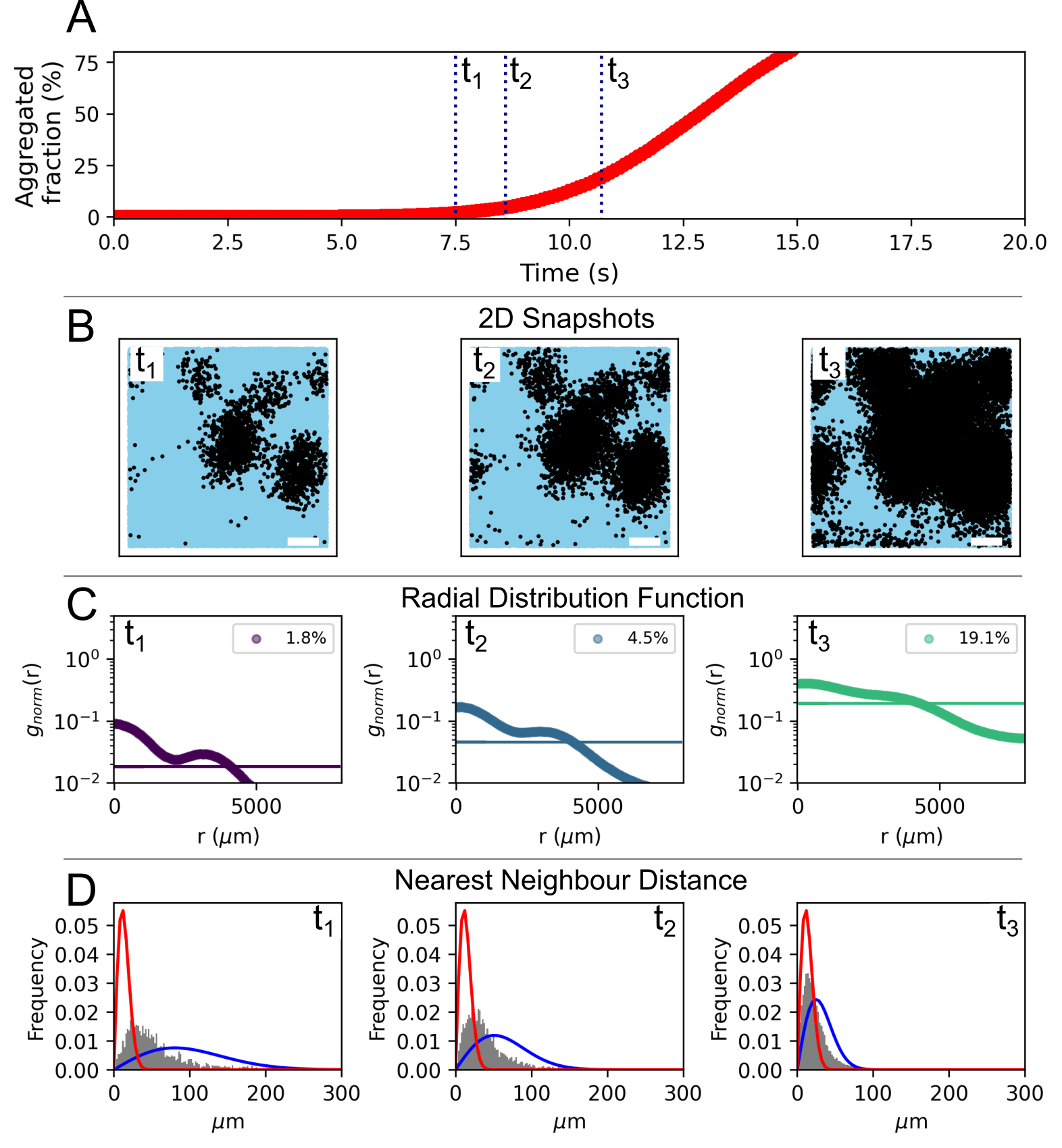}
  \caption{\textbf{Intermediate regime - long range coupling}.}(A) Temporal evolution of total aggregated fraction. The red dots are simulated points. (B) Snapshots of simulations of weak spatial coupling case at fraction aggregated = 2\%, 5\%, 20\%. (C) RDF of (B) (D) Corresponding NND distribution of (B). Scale bar on (B): 1000 \textmu m. Simulation conditions: total cell number = 100000, $k_a$ = 0.00001/s, $k_s$ = 1.0/s, $\sigma$ = 300$\mu m$, coupling kernel: Gaussian, $\Delta t$ = 0.1 s, $\rho$ = 1341.76/$mm^2$, initial aggregated cell number = 0.
  \label{fig:intermediate_large_sigma}
\end{figure}

\subsection{Intermediate regime: switch fraction}
Having considered the limiting cases, we now move on to discussing a system in which both cell-autonomous and cell-to-cell triggers drive the progression of the aggregation pattern. This will be the case in many real systems: there is a non-zero probability of a cell switching state spontaneously, as evidenced by the fact that most neurodegnerative diseases are sporadic, with the exception of prion disease~\cite{PRUSINER1982} and some rare cases of AD~\cite{Banerjee2024,Kovacs2016} where the introduction of infectious material seems to play a role. Additionally, there is also strong evidence that pathology spreads between cells, supporting the presence of cell-to-cell triggers~\cite{Clavaguera2009, Weickenmeier2018}.

To visualize these competing dynamics, we first simulated the intermediate regime with short-range coupling. The simulations reveal that pathology begins with multiple, spatially random cell-autonomous events that then immediately trigger the growth of distinct local clusters, as illustrated in the snapshots in Fig.~\ref{fig:intermediate_regime}B. As the simulation progresses, these initial clusters expand and begin to merge, creating complex spatial patterns. The immediate impact of the spatial coupling can be seen in the RDF (Fig. \ref{fig:intermediate_regime}C). Unlike the purely spontaneous aggregation case ($k_s=0$, Fig.\ref{fig:primary}C), the RDF here exhibits a distinct peak at short distances even at the earliest time point, $t_1$, indicating that the cell-to-cell coupling can affect tissue scale patterning even at early times. As the aggregation fraction increases, this peak grows and broadens, reflecting the expansion and merger of these clusters.

We then extended this analysis to the long-range coupling limit to demonstrate how the characteristic length scale of interactions influences pattern formation. In this regime, the aggregation process is significantly accelerated (Fig. \ref{fig:intermediate_large_sigma}A). The resulting clusters are larger, more diffuse, and merge more rapidly than in the short-range case (Fig. \ref{fig:intermediate_large_sigma}B). This change in morphology is reflected in the RDF, which shows a broader correlation peak extending over larger distances (Fig. \ref{fig:intermediate_large_sigma}C). Together, these simulations illustrate that the interplay between the cell-autonomous rate ($k_a$) and the spatial coupling range ($\sigma$) creates a rich diversity of aggregation patterns.

To quantify the competition between cell-autonomous and cell-to-cell triggers, we compare their respective contribution to the rates. The rate of the cell-autonomous process, $r_a$, is determined by the rate constant $k_a$ and the fraction of healthy cells available to be converted, $(1-n)$: 
$$r_a=(1-n)k_a.$$
The rate of spatial coupling is governed by Eq.~\eqref{pde_coupling}:
$$\frac{\partial n}{\partial t}=k_s\rho(1-n)n+\frac{1}{2}k_s\rho(1-n)\sigma^2\nabla^2 n.$$
The first term represents the local increase in aggregated cells due to cell-to-cell triggers. The second term accounts for effects resulting from a spatially non-uniform distribution of aggregated cells and can be omitted since, before the switch, the cell-autonomous process dominates, and the $n$ remains uniform across space (within the continuum approximation). Hence, the rate of spatial coupling $r_s$ is then expressed as
$$r_s=k_s\rho(1-n)n.$$
This functional form is the hallmark of logistic growth. At the onset of the coupling-dominated phase when $n$ is small, this process is exponential-like, which naturally accounts for the rapid acceleration of disease seen clinically. The growth then slows as the pool of healthy cells is depleted, leading to the characteristic sigmoidal progression curves~\cite{Jack2010}.

The transition or `switch' from cell-autonomous processes to cell-to-cell coupling driving aggregation occurs when the rates $r_s$ and $r_a$ are equal. This condition defines the critical fraction $n_c$:  
$$n_c=\frac{k_a}{k_s}\frac{1}{\rho}.$$
Before the critical fraction $n_c$ is reached ($n<n_c$), the cell-autonomous process dominates. In this regime, $r_a>r_s$ , and the system's dynamics are primarily driven by the creation of new clusters of cells with aggregates.  As the system evolves and $n$ approaches $n_c$, the rates become comparable. Beyond this critical fraction ($n>n_c$), the cell-to-cell mechanism overtakes the cell-autonomous process as the dominant process ($r_s>r_a$), and the growth is largely driven by the expansion of existing clusters rather than the creation of new ones. In this way, the switch fraction $n_c$ provides a quantitative framework for linking the two mechanisms and understanding the behaviour of the system as it transitions from a cell-autonomous phase to a coupling-dominated phase. Interestingly, a mathematically similar transition occurs in the aggregation of proteins within a single homogeneous system, where early on new aggregates are predominantly formed by primary nucleation processes, but once a sufficient concentration of aggregates has accumulated, aggregate self-replication, via secondary nucleation or fragmentation, instead dominates the formation of new aggregates.

\begin{figure}
  \includegraphics[width=1.0\linewidth]{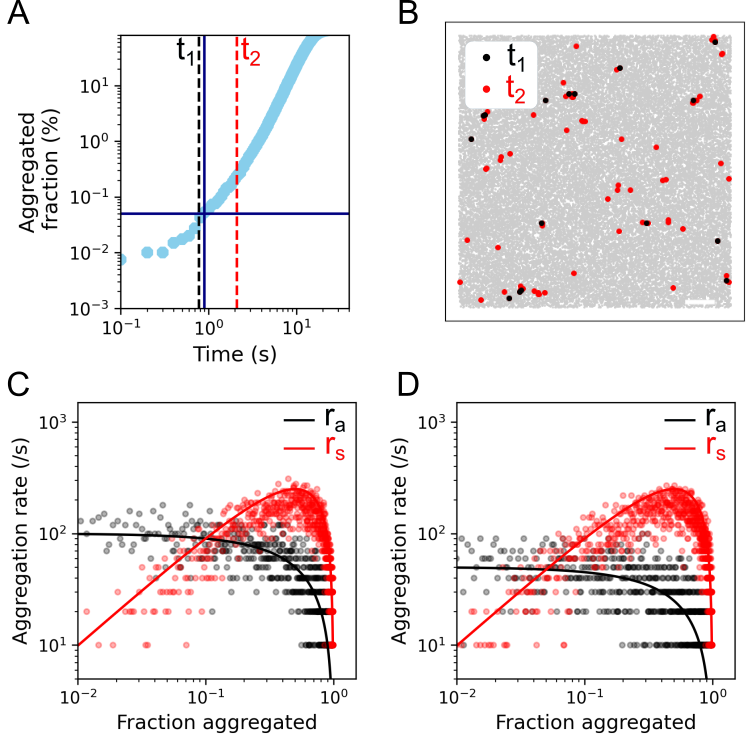}
  \caption{\textbf{Switch fraction of different conditions}.}(A) Selected time points before ($t_1$, red dash line) and after ($t_2$, black dash line) the switch fraction is reached (navy solid lines: switch fraction and switch time).  (B) Distribution of aggregated cells before (black dots) and after (red dots) the switch fraction is reached. Two time points corresponded to $t_1$, and $t_2$ in Fig.~\ref{fig:switch_fraction}A (C, D) Rates of cell-autonomous process ($r_a$) and spatial coupling ($r_s$) versus total fraction aggregated at different rate constants. Simulation conditions: (A,B) total cell number = 39999, $k_a$ = 0.0005/s, $k_s$ = 1.0/s, $\sigma$ = 40$\mu m$, coupling kernel: Gaussian, $\Delta t$ = 0.1 s, $\rho$ = 1341.76/$mm^2$, initial aggregated cell number = 0. (C) total cell number = 10002, $k_a$ = 0.01/s, $k_s$ = 0.1/s, $\sigma$ = 40$\mu m$, coupling kernel: Gaussian, $\Delta t$ = 0.1 s, $\rho$ = 1341.76/$mm^2$, initial aggregated cell number = 0. (D) total cell number = 10002, $k_a$ = 0.005/s, $k_s$ = 0.1/s, $\sigma$ = 40$\mu m$, coupling kernel: Gaussian, $\Delta t$ = 0.1 s, $\rho$ = 1341.76/$mm^2$, initial aggregated cell number = 0.
  \label{fig:switch_fraction}
\end{figure}

Our simulations provide definitive validation for this theoretical prediction of a mechanistic switch, as shown in Fig. \ref{fig:switch_fraction}. The plots of the instantaneous rates versus the fraction aggregated (Fig. \ref{fig:switch_fraction}C, D) clearly demonstrate the crossover at the predicted value of $n_c$ for different kinetic parameters. This crossover visibly marks the transition from a cell-autonomous ($r_a>r_s$) to a growth-dominated ($r_s>r_a$) regime. Furthermore, Fig. \ref{fig:switch_fraction}B offers a  visual illustration of this switch, contrasting the sparse, randomly seeded pattern of aggregates present before the switch ($t_1$) with the denser, cluster-driven pattern that emerges after the switch ($t_2$). Taken together, the results in Figure \ref{fig:switch_fraction} provide a robust validation of our analytical framework, establishing a clear and quantitative link between the microscopic kinetic parameters and the emergent macroscopic dynamics of the system. 

\section{Discussion}

Our multiscale model provides a direct bridge between the microscopic drivers of protein aggregation and the macroscopic patterns of neurodegenerative disease. We show that this type of model exhibits a critical `switch fraction' ($n_c$) - a quantitative metric that predicts when pathology transitions from a cell-autonomous to a coupling-dominated phase. This finding has clear implications for clinical strategy, as it provides a physical basis for stratifying diseases once early biomarkers become more established: conditions dominated by cell-autonomous processes ($k_a\gg k_s$) would be prime candidates for therapies stabilizing the native proteome~\cite{Bulic2010, Bourdenx2021} , whereas those dominated by cell-to-cell propagation ($k_s\gg k_a$) may also be susceptible to immunotherapies targeting seeds~\cite{Yadollahikhales2023,Cummings2023,Vaz2020}. 

A key insight from our model is that it naturally recapitulates the characteristic temporal progression of diseases such as Progressive Supranuclear Palsy~\cite{Jack2010,Huang2024}. The initial, slow accumulation of pathology corresponds to the cell-autonomous regime ($n<n_c$), where the rate of disease progression is limited by the slow, stochastic formation of new, independent seeds. However, once the density of aggregated cells surpasses the switch fraction, the system transitions to the propagation-dominated regime ($n>n_c$). In this phase, the process becomes self-catalyzing, leading to a rapid increase in the number of affected cells as pathology spreads efficiently from existing clusters. This mechanistic switch provides a direct physical explanation for the long prodromal period followed by an aggressive acceleration of pathology that is a hallmark of the clinical presentation of many neurodegenerative disorders. 

From a physical sciences perspective, our work explains how distinct pathological patterns emerge from a single set of underlying microscopic rules. By demonstrating that the balance between cell-autonomous triggers and cell-to-cell propagation governs the outcome, our model reveals three distinct macroscopic regimes: random distributions, the formation of dense clusters with sharp boundaries, and diffuse travelling waves consistent with reaction-diffusion systems described by the Fisher-KPP equation. This work therefore provides a unified framework to interpret the diverse tissue pathologies seen in patients. 

While we have showcased the model here for two-dimensional plane with uniform cell density, it can be extended to 3D and populated with patient-specific anatomical maps to make personalized predictions~\cite{Chung2013,Renier2014,Huang2024}. Furthermore, by incorporating terms for cell death~\cite{Jellinger2001} and aggregate clearance~\cite{Meisl_clearance,Takalo2013}, our model could directly probe the causal links between aggregate load and the rate of neurodegeneration, exploring how different cell types contribute to pathology. This framework can thus be used to guide experimental design, predicting the optimal measurements to reliably distinguish between autonomous and propagating mechanisms in vivo.

\begin{acknowledgments}
S.-H.H. was supported by a Taiwan Cambridge Scholarship.  The work was also supported by the UK Dementia Research Institute (which receives its funding from UK DRI Ltd) and the Royal Society (D.K.) and a UK DRI pilot award (G.M., M.W.C.).
\end{acknowledgments}
\appendix
\section{Total cell fraction aggregated in weak spatial coupling case}
\label{(SI)primary_frac_aggregated}
Here, we assume that there is no coupling between cells, i.e. $k_s\rightarrow 0$. Suppose that the probability of one cell to be triggered by cell-autonomous process in the time step $\Delta t$ is $p_{ca}$. Since $p_{ca}$ is independent of space, we can write down a Markov-like representation of average nucleated cell numbers. $R_{m}=R_{m-1}(1-p_{ca})$ where $R_m$ is the number of nucleated cells within the $m$-th time step (each time step is $\Delta t$). 

Suppose at time step $m=1$, the number of nucleated cells $R_{1}=Np_{ca}$, where $N$ is the total number of grids/cells. Then, we can calculate the total nucleated cells up to time step $m=k$ as 

$$
\Sigma_{m=1}^kR_{m}=N(1-(1-p_{ca})^k)
$$

Taking the limit of $\Delta t \rightarrow 0$ for a step and writing $k = \frac{t}{\Delta t}$ give $\lim_{\Delta t\rightarrow 0} N(1-(1-p_{ca})^{\frac{t}{\Delta t}})=N (1-e^{-k_{a}t})$,

where $k_a$ is the rate of cell-autonomous process. We can also work out the relation between bulk rate of cell-autonomous process and probability of cell-autonomous process in the discrete simulation case: $k_a=-\frac{1}{\Delta t}\log(1-p_{ca})$.

Hence, the percentage of nucleated cells up to time $t$ is 

$$
\frac{N(1-e^{-k_{a}t})}{N}=1-e^{-k_{a}t}
\equiv n(t)
$$

\section{Derivation of the RDF in the spatially uncorrelated limit}
\label{Apdix:RDF_random_case}
In the limiting case of purely cell-autonomous aggregation $(k_s\rightarrow0)$, the conversion of any given cell is an independent event, resulting in a spatially random distribution of aggregated cells. The normalised Radial Distribution Function is defined as the ratio $g_{norm}=g_{agg}(r)/g_{nuc}(r)$, where $g_{agg}(r)$ and $g_{nuc}(r)$ are the RDFs of the aggregated and total cell populations, respectively. 

Because the aggregated cells form a random subset of the total cell population, their spatial distribution is statistically identical to that of the total population. Consequently, the functional form of $g_{agg}(r)$ and $g_{nuc}(r)$ are identical, differing only by a scaling factor proportional to their respective average densities ($\rho_{agg}$ and $\rho_{nuc}$). When the ratio is taken, this functional dependence on distance, $r$, cancels out, leaving only the ratio of the average densities: $$g_{norm}=\frac{\rho_{agg}}{\rho_{nuc}}$$, which is exactly the total fraction of aggregated cells. 

\section{NND distribution in weak spatial coupling case}
\label{(Apdix)primary_NND}
To calculate the NND distribution of aggregate, we are essentially asking the probability distribution of finding the nearest neighbour aggregate at a distance from a reference point. Let us assume that the reference point is at the origin. 

To calculate the probability density of having nearest-neighbour aggregate at a certain distance, we can break this down into two parts. The first part is the probability density not finding an aggregate in the circle with radius $r$, where mathematically speaking is $P(X=0\ not\ in\ radius\ r)$.

Since the events of cell aggregation are independent and random in space, the probability of encountering $k$ aggregation events in the circle follows a Poisson distribution. The probability mass function reads $P(X=k)=\frac{\lambda^ke^{-\lambda}}{k!}$. And the cumulative distribution function reads $P(X\le k)=e^{-\lambda}\Sigma_{j=0}^{\lfloor k \rfloor}\frac{\lambda^j}{j!}$, where $k$ is the number of aggregate events, and $\lambda$ is the expected value of the aggregation events $X$, which is $\pi r^2D$ in the current case. 

Hence, $P(X=0\ not\ in\ radius\ r)$ is equivalent to $P(X\le 1|\lambda=\pi r^2D)$, which is the cumulative distribution function of Poisson distribution of finding less or equal than one event.  By definition, $P(X\le 1|\lambda=\pi r^2D) =(1+\lambda) e^{-\pi r^2D}\propto e^{-\pi r^2D}.$

The second part is the probability density of finding an aggregate in the ring $r\rightarrow r+dr$, which mathematically speaking is $P(X=1\ in\ ring\ r\rightarrow r+dr)$, or equivalently $P(X=1|\lambda=2\pi r Ddr)$. Again, refer to the definition of Poisson distribution, $P(X=1|\lambda=2\pi r Ddr) =  (2\pi rDdr)e^{-2\pi rDdr}\approx 2\pi rDdr.$ Here we make an assumption of small ring width, i.e. $dr\rightarrow 0$. 

Finally, the NND distribution when the aggregates are randomly distributed is 
\begin{equation}
\begin{aligned}
    P&(nearest\ neighbor\ distance\ is\ r\rightarrow r+dr) \\
    \hspace{1cm}&=P(X\le 1|\lambda=\pi r^2D)P(X=1|\lambda=2\pi r Ddr)\\
    &=2\pi rDe^{-\pi r^2D}dr.
\end{aligned}
\end{equation}

\emph{Time step condition for spatial coupling simulation}
To avoid discretisation issues in time, we ensure that $k_s\times\Delta t<0.1$.

\bibliography{apssamp}

\clearpage
\newpage

\onecolumngrid
\begin{center}
  \textbf{\Large Supplementary Materials for: \\[0.2cm] "A cell-level protein aggregation model to predict neurodegenerative disease phenomena"}
\end{center}

\setcounter{equation}{0}
\setcounter{figure}{0}
\setcounter{table}{0}
\setcounter{section}{0}
\setcounter{page}{1} 

\renewcommand{\theequation}{S\arabic{equation}}
\renewcommand{\thefigure}{S\arabic{figure}}
\renewcommand{\thetable}{S\Roman{table}} 
\renewcommand{\thesection}{S\arabic{section}} 


\section{Time Convergence Test}
\begin{figure*}[h!]
  \includegraphics[width=1.0\linewidth]{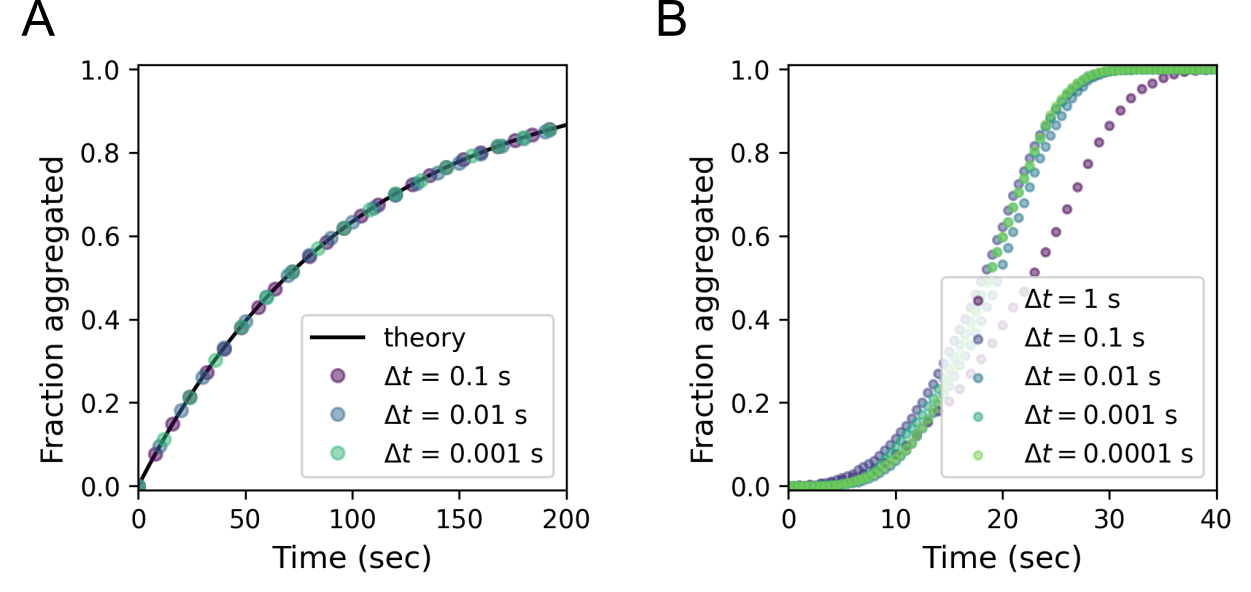}
  \caption{\textbf{Time convergence test.} (A) Average concentration with time at time steps 0.1 sec, 0.01 sec, and 0.001 sec. The solid black line is the analytical theory (Eq.~\ref{eq:prim}). (B) Fraction aggregated with time at time steps 0.1 sec, 0.01 sec, and 0.005 sec. Simulation conditions: (A) total cell number = 100000, $k_a$ = 0.01/s, $k_s$ = 0, $\sigma$ = 200$\mu m$, coupling kernel: Gaussian, $\Delta t$ = 0.1, 0.01, 0.001 s, $\rho$ = 1371.74/$mm^2$, initial aggregated cell number = 1. (B) total cell number = 5000, $k_a$ = 0, $k_s$ = 1.0/s, $\sigma$ = 50$\mu m$, coupling kernel: Gaussian, $\Delta t$ = 1, 0.1, 0.01, 0.001,0.0001 s, $\rho$ = 1371.74/$mm^2$, initial aggregated cell number = 1.}
  \label{fig:SIFig1_time_conv}
\end{figure*}

\end{document}